\documentclass[twocolumn,aps,prb,superscriptaddress,address,amsmath,amssymb,showpacs]{revtex4}
\usepackage[latin9]{inputenc}
\usepackage{amsmath}
\usepackage{amssymb}
\usepackage{graphicx}
\usepackage{esint}
\usepackage[unicode=true,pdfusetitle,
 bookmarks=true,bookmarksnumbered=false,bookmarksopen=false,
 breaklinks=false,pdfborder={0 0 1},backref=false,colorlinks=false]
 {hyperref}
\makeatletter
\@ifundefined{textcolor}{}
{%
 \definecolor{BLACK}{gray}{0}
 \definecolor{WHITE}{gray}{1}
 \definecolor{RED}{rgb}{1,0,0}
 \definecolor{GREEN}{rgb}{0,1,0}
 \definecolor{BLUE}{rgb}{0,0,1}
 \definecolor{CYAN}{cmyk}{1,0,0,0}
 \definecolor{MAGENTA}{cmyk}{0,1,0,0}
 \definecolor{YELLOW}{cmyk}{0,0,1,0}
}

\usepackage{subfigure}
\usepackage{color}

\newcommand{\gtwid}{\mathrel{\raise.3ex\hbox{$>$\kern-.75em\lower1ex\hbox{$\sim$}}}}
\newcommand{\ltwid}{\mathrel{\raise.3ex\hbox{$<$\kern-.75em\lower1ex\hbox{$\sim$}}}}
\makeatother

\begin{document}


\title{Theoretical investigation of the applicability of the Meservey-Tedrow
technique to the surface states of topological insulators}

\author{Matthias G\"otte and Thomas Dahm}

\affiliation{Universit\"at Bielefeld, Fakult\"at f\"ur Physik, Postfach 100131, D-33501
Bielefeld, Germany}

\date{\today}
\begin{abstract}
The spin polarization of topological surface states is of high interest
for possible applications in spintronics. At present, the only technique 
capable to measure the surface state spin texture is spin and angle resolved 
photoemission spectroscopy (SARPES). However, values reported by SARPES
differed strongly. An established technique
to measure the spin polarization of ferromagnetic materials is the so-called
Meservey-Tedrow technique, which is based on spin dependent tunneling from
a superconducting electrode to a ferromagnet. Here, we theoretically investigate how the
Meservey-Tedrow technique can be adapted to topological insulators.
We demonstrate that with a specific device geometry it is possible to
determine the in-plane component
of the spin polarization of topological surface states.
More complex device geometries can access the full momentum dependence of
the spin polarization. We also show that it is possible to extract the
spin-flip scattering rate of surface electrons with the same devices.
\end{abstract}
\maketitle


\section{Introduction}

Since their discovery about a decade ago\cite{Bernevig,Fu,Koenig,Hsieh1,Chen},
topological insulators (TIs) have attracted great interest in the
field of spintronics\cite{Tanaka,Garate,Krueckl,TMR,GJD,Han,Khang,He}. 
This interest originates
from the presence of topologically protected surface states in an
otherwise insulating bulk gap. Particularly, these surface states
have a Dirac cone like dispersion and, due to strong spin orbit coupling,
spin and momentum are locked, i.e. electrons propagating in opposite
directions possess opposite spin. 

Theoretical calculations for three-dimensional TIs like Bi$_{2}$Se$_{3}$
predict that the spin is orthogonal to the momentum and lies mainly in the
surface plane, with a small out-of-plane component in some materials
due to the hexagonal deformation of the Fermi surface for larger wavevectors\cite{Liu:PRB10}.
While this is well confirmed by spin- and angle-resolved photoemission
spectroscopy (SARPES) experiments\cite{Souma}, the measurement
of the degree of spin-polarization is still problematic. While most
theoretical calculations yield values around $50\%-65\%$\cite{Yazyev,Wang,Sanchez},
the values reported by SARPES measurements differ strongly between $\sim45\%$
and $100\%$\cite{Sanchez,Pan:PRL106,Jozwiak,Landolt}. The problem
with this technique is that, depending on the photon energy and the
photon polarization, the spin of the photoelectrons is different
from that of the original electrons in the topological surface states.
\cite{Sanchez,Park} However,
a detailed knowledge about the surface state spin-texture is important,
because it can be crucial for the efficiency of spintronic devices
based on topological insulators.

\begin{figure}
\centering\includegraphics[width=0.9\columnwidth]{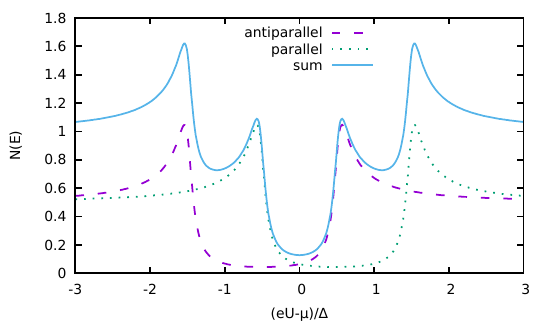}

\caption{\label{fig:Density-of-states}BCS density of states within an applied
magnetic field. The energy of electrons with parallel and antiparallel spin orientation
with respect to the applied magnetic field is shifted in opposite
directions with respect to the chemical potential $\mu$ (dashed purple and
dotted green). In a tunnel junction the total differential conductance is
a superposition of these two spectra depending on the polarization of the other
material. For an unpolarized material one finds an equal superposition (blue),
while for a polarized material the degree of polarization can be obtained from the
asymmetry of the peak heights observed in the differential tunnelling conductance.}
\end{figure}

An established technique to measure the spin polarization of ferromagnetic
materials is a method first used by Meservey and Tedrow
\cite{TedrowMeservey-PRL26,TedrowMeservey-PRB7}. In this technique
spin-dependent tunnelling from a superconductor to a ferromagnet is
used to determine the degree of spin polarization.
Meservey, Tedrow, and Fulde showed that in
thin superconducting aluminum (Al) films the quasiparticle states split in a strong parallel
magnetic field $B$ and the BCS energy spectra of spin-up and spin-down
electrons are thereby shifted by $\pm\mu_{B}B$ with respect to the
original spectrum (see Fig.~\ref{fig:Density-of-states}).\cite{TedrowMeservey-PRL25}
In junctions made out of Al, an insulating barrier (I), and a third material,
this allows to measure the spin polarization of the third material from the conductance
of the junction as the spin of tunnelling electrons is
conserved.\cite{TedrowMeservey-PRL26}
If the third material is a material
in which all spins are oriented either parallel or antiparallel to the
magnetic field, e.g. a ferromagnet, the spectrum is a simple polarization-dependent
superposition of the two shifted spectra like in Fig.~\ref{fig:Density-of-states}.
The polarization can then easily be extracted from the relative height
of the four peaks at $\sim\pm\left(\Delta\pm\mu_{B}B\right)$ in the
spectrum as was shown by Tedrow and Meservey.
\cite{TedrowMeservey-PRL26,TedrowMeservey-PRB7} 

\begin{figure}
\centering\includegraphics[width=0.9\columnwidth]{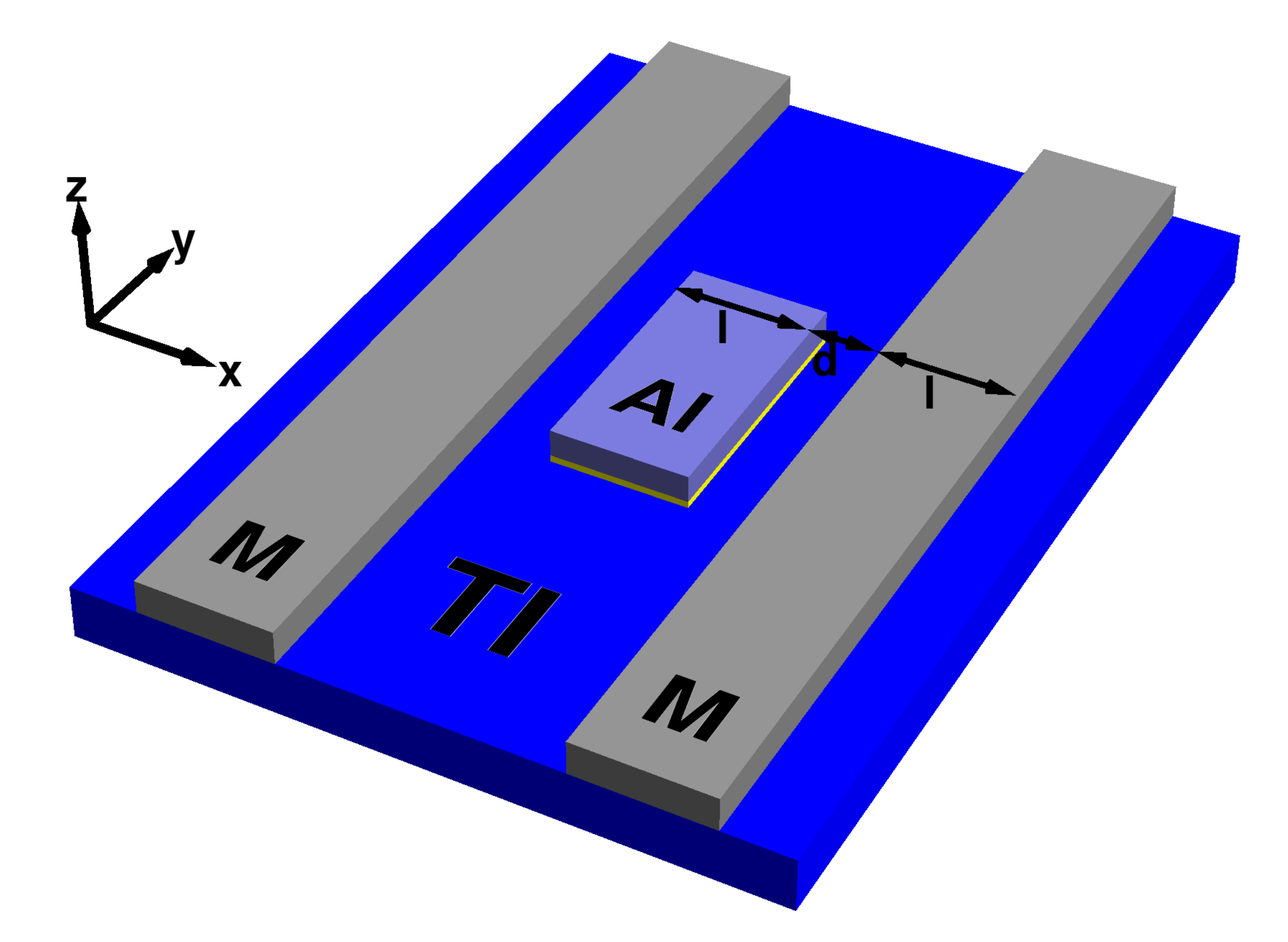}

\caption{\label{fig:device}Basic geometry of a device for measuring the spin polarization
of topological surface states using the Meservey Tedrow technique. The current is injected into
the topological surface states via the aluminum (Al) electrode and
extracted through the metallic electrodes (M). Due to the large spatial
extend of the TI in $y$-direction compared to the distance $d$,
possible scattering processes at the edges of the TI can be neglected.}
\end{figure}

The Meservey-Tedrow technique cannot be applied directly to a topological
insulator surface, because the spin direction rotates around the Fermi
surface. Therefore, the total polarization seen in the tunneling conductance
would average out to zero. In order to still allow a determination of
the spin polarization we suggest to take advantage of the spin-momentum
locking of the topological surface states. The main idea can be understood
from the
device geometry shown in Fig.~\ref{fig:device}. The superconducting Al electrode
(light blue) is placed on top of the topological insulator surface (dark blue)
and a thin insulating barrier (yellow). Two metallic electrodes (grey) are placed
to the left and right of the Al electrode with a distance $d$ between them.
A tunnelling current is fed through the Al electrode and extracted on the
metallic electrodes, which should both be on the same potential. 
Due to the helical spin texture and the Dirac cone like dispersion
of the topological surface states, the mean spin polarization of electrons propagating
in one direction is opposite to that of electrons propagating in the
opposite direction. As a result, the current extracted at the right metallic
electrode will be dominated by one spin direction and the current extracted at 
the left metallic electrode by the opposite spin direction.
From the imbalance of the differential conductance one can then determine
the degree of spin polarization of the topological surface states.

In the following we will provide theoretical calculations of the differential
conductance of such a device and demonstrate that an extraction of the spin
polarization is feasible.\cite{Dissertation} In contrast to most ferromagnetic materials
the density of states of the topological surface states varies strongly
as a function of energy, which needs to be taken into account in the
calculations. Also, the tunnelling conductance will depend on the spin-flip
mean free path of the electrons in the topological surface states. In the
presence of spin-flip scattering an electron that initially propagates in
positive $x$-direction can be scattered and eventually appear at the opposite electrode
in negative $x$-direction. For this reason the distance $d$ between the
electrodes should be chosen smaller than a few spin-flip mean free paths to
be able to extract the spin polarization. Experimental values for the
spin-flip mean free path range between 200~nm in disordered Bi$_2$Se$_3$
\cite{Dufouleur} and a few microns in HgTe \cite{Koenig2}. In section III
we will discuss this dependence on the spin-flip mean free path and show
that it provides the possibility to extract the spin-flip mean free path from
such devices.

\section{Model}

To provide a realistic calculation of the tunnelling conductance of an
Al/I/TI junction, we use a tight binding model for
the Bi$_{2}$Se$_{3}$ class of materials as given in Ref.~\onlinecite{TMR}.
This tight binding model has been derived from bandstructure calculations up to third order in 
momentum $\mathbf{k}$ by Liu et
al\cite{Liu:PRB10} using $\mathbf{k}\cdot\mathbf{p}$ theory.
In $\mathbf{k}$-space the Hamiltonian can be written in terms of
the Dirac $\Gamma$ matrices $\Gamma^{1,2,3,4,5}=\left(\tau_{1}\otimes\sigma_{1},\tau_{1}\otimes\sigma_{2},\tau_{2}\otimes\mathbb{I}_{2\times2},\tau_{3}\otimes\mathbb{I}_{2\times2},\tau_{1}\otimes\sigma_{3}\right)$,
where the Pauli matrices $\tau_{i}$ and $\sigma_{i}$ act in orbital
and spin space, respectively. It takes into account four dominant
bands at the Fermi surface and reads

\begin{equation}
H(\mathbf{k})=\epsilon_{0}(\mathbf{k})\mathbb{I}_{4\times4}+\sum_{i=1}^{4}m_{i}\left(\mathbf{k}\right)\Gamma^{i}+\mathcal{R}_{1}\left(\mathbf{k}\right)\Gamma^{5}+\mathcal{R}_{2}\left(\mathbf{k}\right)\Gamma^{3}.\label{eq:hamiltonian}
\end{equation}
On a bilayer hexagonal lattice the tight-binding parameters can be
defined as follows:\cite{Hao,TMR} 
\begin{eqnarray*}
\epsilon_{0}(\mathbf{k}) & = & C_{0}+2C_{1}\left(1-\cos k_{z}\right)\\
 &  & +\frac{4}{3}C_{2}\left(3-2\cos\frac{1}{2}k_{x}\cos\frac{\sqrt{3}}{2}k_{y}-\cos k_{x}\right)\\
m_{1}(\mathbf{k}) & = & A_{0}\frac{2}{\sqrt{3}}\cos\frac{1}{2}k_{x}\sin\frac{\sqrt{3}}{2}k_{y}\\
m_{2}(\mathbf{k}) & = & -A_{0}\frac{2}{3}\left(\sin\frac{1}{2}k_{x}\cos\frac{\sqrt{3}}{2}k_{y}+\sin k_{x}\right)\\
m_{3}(\mathbf{k}) & = & B_{0}\sin k_{z}\\
m_{4}(\mathbf{k}) & = & M_{0}+2M_{1}\left(1-\cos k_{z}\right)\\
 &  & +\frac{4}{3}M_{2}\left(3-2\cos\frac{1}{2}k_{x}\cos\frac{\sqrt{3}}{2}k_{y}-\cos k_{x}\right)\\
\mathcal{R}_{1}\left(\mathbf{k}\right) & = & 2R_{1}\left(\cos\sqrt{3}k_{y}-\cos k_{x}\right)\sin k_{x}\\
\mathcal{R}_{2}\left(\mathbf{k}\right) & = & \frac{16}{3\sqrt{3}}R_{2}\left(\cos\frac{\sqrt{3}}{2}k_{y}-\cos\frac{3}{2}k_{x}\right)\sin\frac{\sqrt{3}}{2}k_{y}
\end{eqnarray*}
The corresponding model parameters for Bi$_{2}$Se$_{3}$, which we
consider here as an example, are derived from Liu et al\cite{Liu:PRB10}
using the atomic distances $a=4.14\textrm{\AA}$ and $c=\frac{28.64}{15}\textrm{\AA}$\cite{Zhang:APL09}:
$A_{0}=0.804\textrm{eV}$, $B_{0}=1.184\textrm{eV}$, $C_{1}=1.575\textrm{eV}$,
$C_{2}=1.774\textrm{eV}$, $M_{0}=-0.28\textrm{eV}$, $M_{1}=1.882\textrm{eV}$,
$M_{2}=2.596\textrm{eV}$, $R_{1}=0.713\textrm{eV}$, and $R_{2}=-1.597\textrm{eV}$.
The parameter $C_{0}$ adds only a tiny energy shift and will therefore
be neglected in the following.

As regards the aluminum electrode, only its density of states near the Fermi
level is of importance for the tunnelling current.
Thus, for simplicity we describe the superconducting aluminum by 
Hamiltonian Eq.~\eqref{eq:hamiltonian} with only one orbital and parameters $C_{\textrm{Al}}\equiv C_{1}=C_{2}=0.25\textrm{eV}$,
$C_{0}=-0.75\textrm{eV}$ and $A_{0}=B_{0}=M_{0}=M_{1}=M_{2}=R_{1}=R_{2}=0$.
The parameter $C_{0}$ was
chosen such that the center of the Al band fits the Fermi level of
the TI. In this way the normal state density of states of the Al band is
nearly constant within the bulk gap of the TI. The BCS density of states in the superconducting state is described by the Dynes
formula, which accounts for a finite lifetime broadening $\Gamma$ in the aluminum: \cite{Dynes}
\begin{equation}
N_{\pm}\left(E\right)=\Re\frac{\left|E\pm\mu_{B}B\right|-i\Gamma}{\sqrt{\left(\left|E\pm\mu_{B}B\right|-i\Gamma\right)^{2}-\Delta^{2}}}.\label{eq:BCS}
\end{equation}
Here, $E$ is energy, $\mu_{B}$ the electron magnetic moment, $B$
the applied magnetic field, $\Delta=0.35\textrm{meV}$ the superconducting
gap of Al, and $\Gamma=0.03\textrm{meV}$. In order to calculate the transition rate
of the junction, we Fourier-transform the Hamiltonian Eq.~\eqref{eq:hamiltonian}
into real space onto its lattice in $z$-direction, i.e. in the direction
perpendicular to the junction. Periodic boundary conditions are used in the in-plane
directions which allows to keep the in-plane momenta $k_{x}$ and $k_{y}$
as good quantum numbers. 

Using Fermi's golden rule
\begin{equation}
\Gamma_{mn}=\frac{2\pi}{\hbar}\delta\left(E_{n}-E_{m}\right)\left|\left\langle n\left|H_{T}\right|m\right\rangle \right|^{2},
\end{equation}
for the transition rate from an initial state $\left|m\right\rangle $
into a final state $\left|n\right\rangle $, the total tunnelling current
from the Al film into the TI, which takes into account tunnelling processes
at finite temperature in both directions, is given by
\begin{eqnarray}
I\left(U\right) & = & \frac{2\pi e}{\hbar}\sum_{m,n}\left[f\left(E_{m}-eU\right)-f\left(E_{n}\right)\right]\cdot\nonumber \\
 &  & \left|\left\langle n\left|H_{T}\right|m\right\rangle \right|^{2}\delta\left(E_{n}-E_{m}\right)
\end{eqnarray}
for a bias voltage $U$ between the Al electrode and the TI. Here,
$m$ and $n$ number the unperturbed eigenstates of the Al film and the TI, respectively.
The Fermi function
\begin{equation}
f\left(E\right)=\frac{1}{1+e^{\frac{E}{k_{B}T}}}
\end{equation}
describes the occupation of these states at finite temperature. 
The insulating barrier is modeled by a tunneling Hamiltonian of
the form
\begin{equation}
H_{T}=- C_B \sum_{k_x,k_y,\alpha.\sigma} d^\dagger_{k_x,k_y,\alpha.\sigma}
c_{k_x,k_y,\sigma} + \mathrm{h.c.} \, ,\label{eq:thamilton}
\end{equation}
where $d^\dagger_{k_x,k_y,\alpha.\sigma}$ creates an electron in
orbital $\alpha$ with spin $\sigma$ in the top layer
of the topological insulator and $c_{k_x,k_y,\sigma}$
destroys an electron in the bottom layer of the aluminum.

The
differential conductance (DC) which is usually measured in experiments
is obtained from the derivative of $I$ with respect to $U$
\begin{eqnarray}
G\left(U\right)=\frac{dI}{dU} & = & \frac{\pi e^{2}}{2\hbar k_{B}T}\sum_{m,n}\frac{1}{\cosh^{2}\frac{E_{m}-eU}{2k_{B}T}}\cdot\nonumber \\
 &  & \left|\left\langle n\left|H_{T}\right|m\right\rangle \right|^{2}\delta\left(E_{n}-E_{m}\right).\label{eq:dif_con}
\end{eqnarray}

\section{Calculations}

In this section we derive a method to calculate the polarization of
the topological surface states from a given tunnel spectrum. As pointed out
above we cannot simply take the differential conductance of an Al/I/TI junction but have
to exploit the locking between spin and propagation direction.
Concerning our calculations
the propagation direction of an electron can be obtained via its group
velocity $\mathbf{v}\propto\frac{\partial E}{\partial\mathbf{k}}$.
We first consider the device geometry shown in Fig.~\ref{fig:device}.
Other geometries will be discussed in Appendix~\ref{sec:Geometrical-factor}.

In order to obtain useful approximation formulas for the differential
conductance we start from an analytical approximation of the topological 
surface states, which is valid in the vicinity of the Dirac node. The results of this
approximate
calculation will be compared with full numerical results below.
When we expand Hamiltonian Eq.~\eqref{eq:hamiltonian} up to second order in
$\mathbf{k}$, we can derive an analytical expression for the four components
of the surface state wave function:\cite{TMR}
\begin{equation}
\psi_{\pm}\left(p,\varphi\right)=\frac{1}{2}\left(\begin{array}{c}
\pm\sqrt{1+p}e^{-i\left(\varphi-\frac{\pi}{2}\right)}\\
\sqrt{1+p}\\
\mp\sqrt{1-p}e^{-i\left(\varphi-\frac{\pi}{2}\right)}\\
\sqrt{1-p}
\end{array}\right).\label{eq:psiTI}
\end{equation}
Here, $\pm$ is for the upper and lower Dirac cone, $\varphi$ is the in-plane polar angle of the
momentum, $-1\leq p\leq1$ is the degree of spin-polarization of the
surface states, and the orientation of the spin is given by the phase
$e^{-i\left(\varphi-\frac{\pi}{2}\right)}$, i.e. it is always rotated
by $\frac{\pi}{2}$ with respect to $\varphi$. The corresponding eigenenergies
only depend on
the magnitude $k=\sqrt{k_{x}^{2}+k_{y}^{2}}$ of the in-plane momentum\cite{TMR}
\begin{equation}
E_{\pm}=-\frac{C_{1}M_{0}}{M_{1}}+\left(C_{2}-\frac{C_{1}}{M_{1}}M_{2}\right)k^{2}\pm A_{0}\sqrt{1-\frac{C_{1}^{2}}{M_{1}^{2}}}k.
\end{equation}
These surface states describe an isotropic Dirac cone, i.e. they neglect
the hexagonal deformation, which is of third
order in $\mathbf{k}$. The position of the Dirac node is at the
energy $E_{0}=-\frac{C_{1}M_{0}}{M_{1}}=0.234\textrm{eV}$. 

In our model for the Al electrode we simply have two degenerate eigenstates 
for each pair of $k_{x}$ and
$k_{y}$. The spatial dependence in $z$-direction
is a superposition of an incoming and a reflected wave and thus given by
$\sin zk_{z}$. An appropriate linear combination of these eigenstates then 
leads to states with a specific spin polarization. For a spin polarization
within the $x$-$y$ surface plane, this linear combination is given
by
\begin{eqnarray}
\psi_{\textrm{Al}}\left(z,k_{z},\varphi_{\textrm{Al}}\right) & = & \frac{1}{\sqrt{2}}\sin zk_{z}\left(\begin{array}{c}
e^{-i\varphi_{\textrm{Al}}}\\
1
\end{array}\right)\label{eq:psiAl}
\end{eqnarray}
where $\varphi_{\textrm{Al}}$ is the in-plane polar angle of the polarization
with respect to the $k_{x}$-axis and
\begin{eqnarray}
k_{z} & \approx & \arccos\frac{C_{0}+C_{\textrm{Al}}\left(2+k^{2}\right)-E}{2C_{\textrm{Al}}}
\end{eqnarray}
for given momenta $k_{x}$ and $k_{y}$ and energy $E$. Using Eq.
\eqref{eq:psiTI} and \eqref{eq:psiAl} with $z=1$ for the bottom layer of the aluminum, 
the transfer matrix element of the junction can be calculated
\begin{equation}
\left|\left\langle \psi_{\textrm{Al}}\left|H_{T}\right|\psi_{\pm}\right\rangle \right|^{2}=\frac{1}{2}C_{B}^{2}\sin^{2}k_{z}(k)\left[1\mp p\sin\left(\varphi_{\textrm{Al}}-\varphi\right)\right],
\end{equation}
where $C_{B}$ is the hopping matrix element of the barrier. This
can then be inserted into Eq.~\eqref{eq:dif_con} multiplied with
the shifted BCS density of states. To get the DC with respect to the
electrode in positive $x$-direction we follow Ref.~\onlinecite{TMR} and
introduce a function $f\left(\varphi\right)$, which gives the probability
that an electron initially propagating under an angle of $\varphi$ ends
up at this electrode. $f\left(\varphi\right)$ depends on the geometry
of the device and can also be used to include effects like spin scattering in
the TI. For the simple case shown in Fig.~\ref{fig:device}, where
all electrons with a positive group velocity component in $x$ direction
end up at that electrode it is simply\cite{TMR} 
\begin{equation}
f\left(\varphi\right)=\begin{cases}
1 & \textrm{for}\,\varphi\in\left[-\frac{\pi}{2},\frac{\pi}{2}\right]\\
0 & \textrm{else}.
\end{cases}\label{eq:f3D}
\end{equation}
More complex cases are discussed in Appendix~\ref{sec:Geometrical-factor}.
In the following we assume that $f\left(\varphi\right)$ is an even function, 
i.e. $f\left(\varphi\right)=f\left(-\varphi\right)$, which is always
satisfied when mirror symmetry with respect to the $x$-$z$-plane holds.

\begin{figure}
\centering\includegraphics[width=0.9\columnwidth]{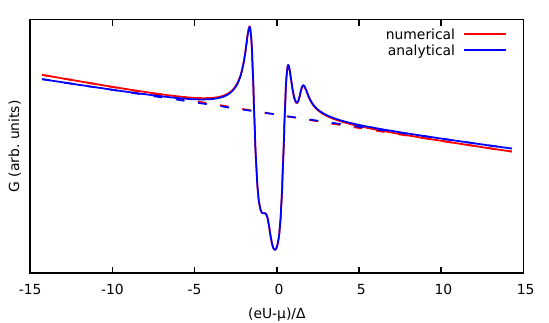}

\caption{\label{fig:Calculated-DC}Calculated DC of the Al/I/Bi$_{2}$Se$_{3}$
junction using $T=0.4\textrm{K}$, $\varphi_{\textrm{Al}}=\frac{\pi}{2}$, $p=1$, and
$\mu=0.2\textrm{eV}$, for the analytical approximation (red) and
the numerical calculation (blue) based on the full Hamiltonian Eq.
\eqref{eq:hamiltonian}. The dashed lines are the correction functions
$h\left(U\right)$, fitted to the outermost tenth on each side of
the DC curves. We used a third order polynomial for the fit.}
\end{figure}

Let us choose the applied magnetic field to point in the direction of
$\varphi_{\textrm{Al}}$.
Then, the differential conductance Eq.~\ref{eq:dif_con} consists of two
contributions coming from the electrons with spin either parallel or
antiparallel to the magnetic field. For the contribution with parallel
spin we find
\begin{eqnarray}
\lefteqn{G_{-}\left(T,U,\varphi_{\textrm{Al}}\right)=}\\
 &  & \frac{\textrm{const.}}{T}\int_{0}^{k_{0}}dkk\int_{-\pi}^{\pi}d\varphi\Bigg(f\left(\varphi\right)\frac{\left|\left\langle \psi_{\textrm{Al}}\left|H_{T}\right|\psi_{+}\right\rangle \right|^{2}}{\cosh^{2}\left(\frac{E_{+}-eU}{2k_{B}T}\right)}N_{-}\left(E_{+}\right)\nonumber \\
 &  & +f\left(\varphi-\pi\right)\frac{\left|\left\langle \psi_{Al}\left|H_{T}\right|\psi_{-}\right\rangle \right|^{2}}{\cosh^{2}\left(\frac{E_{-}-eU}{2k_{B}T}\right)}N_{-}\left(E_{-}\right)\Bigg)\nonumber \\
 & = & \int_{-\pi}^{\pi}d\varphi
 f\left(\varphi\right)\left(1-p\sin\varphi_{\textrm{Al}}\cos\varphi\right)\frac{\textrm{const.}}{T}\int_{0}^{k_{0}}dkk\label{eq:G-2}
 \\
 &  & \cdot\left(\frac{\sin^{2}k_{z}(k)N_{-}\left(E_{+}\right)}{\cosh^{2}\left(\frac{E_{+}-eU}{2k_{B}T}\right)}+\frac{\sin^{2}k_{z}(k)N_{-}\left(E_{-}\right)}{\cosh^{2}\left(\frac{E_{-}-eU}{2k_{B}T}\right)}\right)\nonumber \\
 & = & \int_{-\pi}^{\pi}d\varphi f\left(\varphi\right)\left(1-p\sin\varphi_{\textrm{Al}}\cos\varphi\right)G_{-}'\left(T,U\right).\label{eq:G-}
\end{eqnarray}
Here, the DC becomes a product of a $\varphi$-integral,
which depends on the geometry of the device and the relative polarization
of the Al film with respect to the TI, and the term $G_{-}'\left(T,U\right)$,
which contains the densities of states of the two materials. Analogously, we find for electrons
with spin oriented antiparallel 
\begin{eqnarray}
\lefteqn{G_{+}\left(T,U,\varphi_{\textrm{Al}}\right)=}\label{eq:G+}\\
 &  & \int_{-\pi}^{\pi}d\varphi f\left(\varphi\right)\left(1+p\sin\varphi_{\textrm{Al}}\cos\varphi\right)G_{+}'\left(T,U\right)\nonumber 
\end{eqnarray}
with $G_{+}'\left(T,U\right)$ depending on $N_{+}\left(E\right)$
instead of $N_{-}\left(E\right)$. 
The total DC of the junction is
then
\begin{eqnarray}
\lefteqn{G\left(T,U,\varphi_{\textrm{Al}}\right)=}\\
 &  & G_{-}\left(T,U,\varphi_{\textrm{Al}}\right)+G_{+}\left(T,U,\varphi_{\textrm{Al}}\right)\nonumber \\
 & \overset{\textrm{Eq. \eqref{eq:f3D}}}{=} & \left(\pi-2p\sin\varphi_{\textrm{Al}}\right)G_{-}'\left(T,U\right)\label{eq:DC_ana}\\
 &  & +\left(\pi+2p\sin\varphi_{\textrm{Al}}\right)G_{+}'\left(T,U\right).\nonumber 
\end{eqnarray}
This is shown in Fig.~\ref{fig:Calculated-DC} as a function of bias voltage
$U$ for $T=0.4\textrm{K}$, $B=3\textrm{T}$, $\varphi_{\textrm{Al}}=\frac{\pi}{2}$,
$p=1$, and a chemical potential of $\mu=0.2\textrm{eV}$. 
The chemical potential is chosen
somewhat below the Dirac node, i.e. at an energy where the hexagonal
deformation of the Dirac-cone is small. For comparison, we also show the result
of a numerical calculation of the differential conductance Eq.~\eqref{eq:dif_con}.
This calculation was based on the full Hamiltonian Eq.~\eqref{eq:hamiltonian}
on a hexagonal lattice with $50$ layers along $z$ ($\left[001\right]$). 
The momenta $k_{x}$ and $k_{y}$ were uniformly distributed
over the first Brillouin zone with a discretization of $\frac{2}{\sqrt{3}}\frac{2\pi}{N}$
and $N=48000$, corresponding to a sample width of about $20\text{\ensuremath{\mu}m}$.
In spite of the simplifications made in the analytical approximations,
there is only a small deviation of the two DC curves. Please note that
one can discern four peaks in the DC curves, even though the surface states
are fully polarized here ($p=1$). The physical reason for this is the spin
texture of the topological surface states: even though the current from the
Al electrode to the M electrode in positive $x$-direction is dominated by
spin up electrons, there is a small but finite probability that a spin down
electron can tunnel into a topological surface state with positive group
velocity $v_x$ in $x$-direction. 

In the following we demonstrate that the polarization $p$ of the topological 
surface states can be reliably inferred from such differential conductance 
curves. In Ref.~\onlinecite{TedrowMeservey-PRB7} Tedrow and Meservey
derived a formula to extract the polarization of ferromagnets from the
four peak heights of the DC curve. Here, we adapt that formula
to the present case and show that the polarization $p$ can be obtained
from it within a good approximation.
If the density of states of the TI and Al film were constant as
a function of energy except for the BCS density of states, we could
express $G_{-}'\left(T,U\right)$ and $G_{+}'\left(T,U\right)$ in
terms of the unsplit DC $G'\left(T,U\right)$.\cite{TedrowMeservey-PRB7}
Defining $F_{\pm}=\int_{-\pi}^{\pi}d\varphi f\left(\varphi\right)\left(1\pm
  p\sin\varphi_{\textrm{Al}}\cos\varphi\right)$
the differential conductance $g$ at some arbitrary bias voltage $x=U-\frac{\mu}{e}$ could be
written
\begin{equation}
g(x) =  F_{+}G'\left(T,x+b\right)+F_{-}G'\left(T,x-b\right),
\label{eq:gx}
\end{equation}
where $b\sim\frac{\mu_{B}B}{e}$ is the splitting of spin up and spin
down densities of states. One now evaluates $g(x)$ at four bias voltages
$\pm x$ and $\pm(x-2b)$, where $x \sim\frac{\Delta+\mu_{B}B}{e}$ is chosen
at the peak position of the outermost peak.
The values of the four conductances $g_{1}$ to $g_{4}$ (from left to right)
are then given as
\begin{eqnarray}
g_{1} & = & F_{+}G'\left(T,-x+b\right)+F_{-}G'\left(T,-x-b\right)\label{eq:peak1}\\
g_{2} & = & F_{+}G'\left(T,-x+3b\right)+F_{-}G'\left(T,-x+b\right)\\
g_{3} & = & F_{+}G'\left(T,x-b\right)+F_{-}G'\left(T,x-3b\right)\\
g_{4} & = & F_{+}G'\left(T,x+b\right)+F_{-}G'\left(T,x-b\right).\label{eq:peak4}
\end{eqnarray}
Assuming that $G'\left(T,x\right)$ is a symmetric function of $x$, this set of
equations can be solved and the polarization $p$ can be obtained from the four conductances
$g_{1}$ to $g_{4}$ leading to the formula
\begin{equation}
p=\frac{\left(g_{4}-g_{2}\right)-\left(g_{1}-g_{3}\right)}{\left(g_{4}-g_{2}\right)+\left(g_{1}-g_{3}\right)}\frac{\gamma}{\sin\varphi_{\textrm{Al}}}.\label{eq:pol}
\end{equation}
Here, the factor 
\begin{equation}
\gamma=\frac{\int_{-\pi}^{\pi}d\varphi f\left(\varphi\right)}{\int_{-\pi}^{\pi}d\varphi f\left(\varphi\right)\cos\varphi}\label{eq:gamma}
\end{equation}
accounts for the geometry of the device. For $f\left(\varphi\right)$
given by Eq.~\eqref{eq:f3D} we have $\gamma=\frac{\pi}{2}$. 

Since
we required that we can write the splitted DC curves Eq.~\eqref{eq:G-}
and \eqref{eq:G+} in terms of the unsplit curve $G'\left(T,U\right)$
in the derivation of Eq.~\eqref{eq:pol}, Eq.~\eqref{eq:pol} becomes
inaccurate if the densities of states of spin-up and
spin-down electrons in the superconductor are not the same function
of energy. This is the case if spin-orbit scattering in the superconductor
is present. However, concerning thin aluminum films, spin-orbit scattering
is small and the deviation from Eq.~\eqref{eq:pol} should therefore
be negligible.\cite{TedrowMeservey-PRB7} In the case of significant
spin-orbit scattering in the superconductor, one can still fit the
DC curves to obtain the polarization, if the energy dependence of
the separate spin densities of states are known.

In the above calculation we assumed the normal state densities of states to be
constant as a function of energy. Since this is not the case for TIs
we have to remove this energy dependence from the measured DC in order
to calculate the polarization $p$. This can be done by fitting the
DC curve with a low order polynomial $h\left(U\right)$, after cutting
out the part with a high influence of the BCS density of states, as
shown in Fig.~\ref{fig:Calculated-DC} (dashed line). Afterwards one then multiplies 
$G\left(T,U,\varphi_{\textrm{Al}}\right)$ with $\frac{1}{h\left(U\right)}$ and
analyzes this corrected DC curve.
\begin{figure}
\centering\includegraphics[width=0.9\columnwidth]{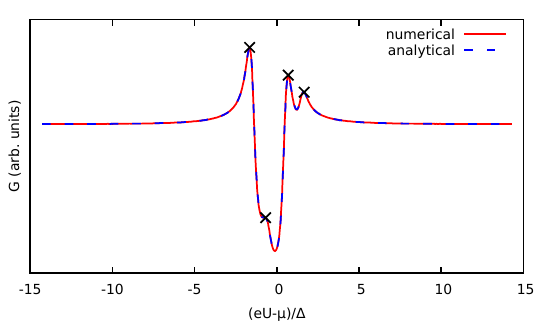}

\caption{\label{fig:DC-corrected}DC with respect to the metallic electrode
in positive $x$-direction of the Al/I/Bi$_{2}$Se$_{3}$ junction
shown in Fig.~\ref{fig:Calculated-DC} after multiplying with $\frac{1}{h\left(U\right)}$,
for the analytical approximation (red, solid) and the numerical calculation
based on the full Hamiltonian Eq.~\eqref{eq:hamiltonian} (blue,
dotted).}
\end{figure}
The corrected DC curves are shown in Fig.~\ref{fig:DC-corrected}.
While there was a small difference between our analytical and numerical
calculations in Fig.~\ref{fig:Calculated-DC}, the curves now agree
very well and thereby support the general validity of this procedure
and of Eq.~\eqref{eq:pol}. Because of the broadening of the BCS
density of states, the exact peak maxima in general do not fulfill
the symmetry requirement around $\mu$. To reduce the error on the
polarization value, the positions for the $g_{i}$ have to be chosen
such that the slope of the DC at these positions is small. Here, we
choose them such that the largest outer and largest inner peak, which
are either $g_{1}$ and $g_{3}$ or $g_{4}$ and $g_{2}$, are met
exactly. The other two positions are then automatically given by the
symmetry requirement around $\mu$. Applying this to the analytical
DC curve in Fig.~\ref{fig:DC-corrected} (black crosses) yields $p\approx0.9996$, 
consistent with the model parameter of $p=1$. 
The deviation from the absolute value of the actual
polarization of $100\%$ is only $0.04$ percentage points.
It is however the inverse of the actual polarization for $k_{x}<0$ at the
lower Dirac cone. The numerical
DC curve yields $p\approx1.0038$, with only a slightly larger deviation.
Applying the same formula, with the same $\gamma$ factor, to the DC of the 
opposite metallic electrode, we find $p\approx-1.0038$,
i.e. the same value with opposite sign. 

\begin{figure}
\centering\includegraphics[width=0.9\columnwidth]{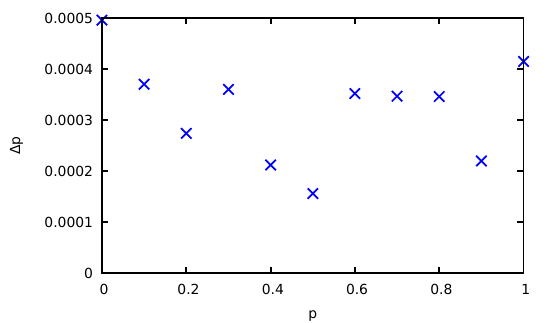}

\caption{\label{fig:Delta_p}Absolute difference $\Delta{p}$ of 
extracted and actual polarization using $T=0.4\textrm{K}$, $\varphi_{\textrm{Al}}=\frac{\pi}{2}$ 
and various polarization values $p$. The difference is always of the same order of magnitude 
with no systematic polarization dependence.}
\end{figure}
Equation
\eqref{eq:pol} is based on the topological surface states given by
Eq.~\eqref{eq:psiTI} where the spin is always perpendicular to the
in-plane momentum. As this may not always be the case it is useful
to rewrite Eq.~\eqref{eq:pol} such that it does not depend on the
direction of the magnetic field but instead on the angular difference $\Delta\varphi_{\textrm{Al}}$
of magnetic field and surface state polarization. When we account for the
counterclockwise rotation of the spin in the lower Dirac cone as well, 
the new model independent formula reads
\begin{equation}
p=\frac{\left(g_{1}-g_{3}\right)-\left(g_{4}-g_{2}\right)}{\left(g_{1}-g_{3}\right)+\left(g_{4}-g_{2}\right)}\frac{\gamma}{\cos\Delta\varphi_{\textrm{Al}}}.\label{eq:pol_dphi}
\end{equation}
Provided that the $\gamma$ factor is calculated for a function $f\left(\varphi\right)$ that 
gives the probability with respect to the electrode in positive direction, 
it always yields the correct polarization value for surface electrons propagating to the considered electrode.
In practice the magnetic field should be oriented parallel to the
polarization of the TI ($\Delta\varphi_{\textrm{Al}}=0$), i.e. such that the
DC becomes maximal. By this, on the one hand we get the orientation
of the surface state spin and on the other hand maximize the tunneling
current. For $\gamma=1$, which is found for $f\left(\varphi\right)=\delta\left(\varphi\right)$,
Eq.~\eqref{eq:pol_dphi} coincides with the formula of Tedrow and
Meservey\cite{TedrowMeservey-PRB7}.

When we apply the above scheme to analytical DC curves for various polarization values, 
the difference $\Delta{p}=p_{\textrm{ex}}-p$ between extracted polarization and actual 
polarization is always of the same order of magnitude, as shown in Fig.~\ref{fig:Delta_p}. 
There also seems to be no correlation between $\Delta{p}$ and $p$.

\begin{figure}
\centering\includegraphics[width=0.9\columnwidth]{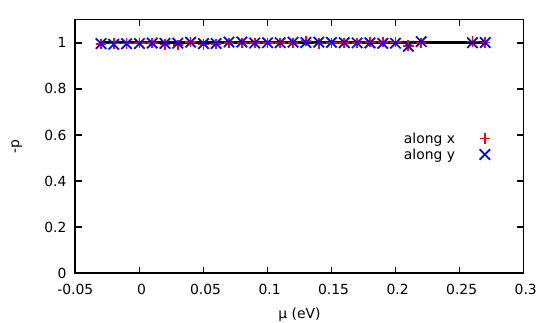}

\caption{\label{fig:p(mu)}Extracted polarization as a function of $\mu$ for
$B=3\textrm{T}$ and $T=0.4\textrm{K}$ (crosses). The measured polarization
shows no sign of the $\mu$ dependent hexagonal out of plane tilt
of the spin polarization, as there is no systematic deviation from
$p=1$ (black line) for both spatial directions. Values near the Dirac
node are missing because the minimum in the density of states makes
it difficult to get a correct fit.}
\end{figure}

In Fig.~\ref{fig:p(mu)} we show the polarization extracted from DCs of the full Hamiltonian as a function of
chemical potential $\mu$. We see no systematic deviation from $p=-1$ .
This is not intuitively clear considering the energy dependent hexagonal
deformation of the Fermi surface along with an out of plane tilt of
the spin polarization. To analyze the influence of a hexagonal deformation
we therefore assume a correction of the surface
state approximation, where the spin is tilted out of the surface plane
with alternating sign with a threefold period: 
\begin{equation}
\psi_{\pm}^{\prime}\left(p,q,\varphi\right)=\frac{1}{2}\left(\begin{array}{c}
\pm\sqrt{1\pm q\cos3\varphi}\sqrt{1+p}e^{-i\left(\varphi-\frac{\pi}{2}\right)}\\
\sqrt{1\mp q\cos3\varphi}\sqrt{1+p}\\
\mp\sqrt{1\pm q\cos3\varphi}\sqrt{1-p}e^{-i\left(\varphi-\frac{\pi}{2}\right)}\\
\sqrt{1\mp q\cos3\varphi}\sqrt{1-p}
\end{array}\right)\,.\label{eq:psi_TI_pq}
\end{equation}
Here, $0\le q\le1$ is the absolute value of the out of plane polarization.
The spin expectation values of this improved approximation are given by
\begin{eqnarray}
n_{x}=\left\langle
  \psi_{\pm}^{\prime}\left|\Sigma_{x}\right|\psi_{\pm}^{\prime}\right\rangle
& = & \pm\sqrt{1-q^{2}\cos^{2}3\varphi}\; p\sin\varphi\\
n_{y}=\left\langle
  \psi_{\pm}^{\prime}\left|\Sigma_{y}\right|\psi_{\pm}^{\prime}\right\rangle
& = & \mp\sqrt{1-q^{2}\cos^{2}3\varphi}\; p\cos\varphi\\
n_{z}=\left\langle \psi_{\pm}^{\prime}\left|\Sigma_{z}\right|\psi_{\pm}^{\prime}\right\rangle  & = & \pm q\cos3\varphi
\end{eqnarray}
With this new approximation, the ratio of the DC peaks remains
basically the same, with a new factor
\begin{equation}
\gamma^{\prime}=\frac{\int_{-\pi}^{\pi}d\varphi f\left(\varphi\right)}{\int_{-\pi}^{\pi}d\varphi f\left(\varphi\right)\sqrt{1-q^{2}\cos^{2}3\varphi}\cos\varphi}\label{eq:gamma_prime}
\end{equation}
which depends on $q$. So it becomes possible to calculate the deviation
from Eq.~(\ref{eq:pol_dphi}) for specified values of $q$ and estimate possible
errors. For the case shown in Fig.~\ref{fig:p(mu)} a maximal out-of-plane
polarization of $q\approx0.17$ at the lower edge of the bulk gap
yields a relative deviation $\frac{\gamma^{\prime}}{\gamma}\approx1.008$
for DC measurements along the $x$-direction. This is only of the order
of the measurement accuracy. Along the $y$-direction (replace $\cos3\varphi$
with $\cos\left(3\varphi-\frac{\pi}{2}\right)$), this ratio is
even slightly smaller $\frac{\gamma^{\prime}}{\gamma}\approx1.007$,
because the out-of-plane polarization is zero along the $y$-axis
and states close to the axis contribute strongest to the DC. 
In the extreme case of $f\left(\varphi\right)=\delta\left(\varphi\right)$,
which would be valid for a two dimensional device,
these deviations become somewhat larger, but are still small. At the
lower end of the bulk gap, a deviation of $\frac{\gamma^{\prime}}{\gamma}\approx1.015$
could be expected. Note, however that in this case, the out-of-plane
polarization strongly depends on the measurement direction. Along
some crystal axes it reaches the maximum value, while for others it
completely vanishes. So, by varying the measurement direction, one
can get rid of the out-of-plane spin component in order to access
the in-plane component. It is however unlikely that the out-of-plane
component can be determined from how $\gamma$ varies as a function
of $\varphi$. A device that is capable of measuring along certain
crystal axes is presented in appendix \ref{sub:U-shaped-geometry}.

\begin{figure}
\centering\includegraphics[width=0.9\columnwidth]{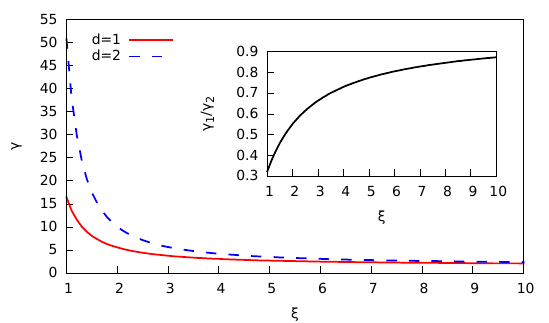}

\caption{\label{fig:gamma_spin-flip}Geometrical factor $\gamma$ with spin-flip
scattering as a function of spin-flip mean free path for two values
of distance $d=1$ and $d=2$. Here, the width of the electrodes is $l=1$.
The inset shows the ratio of the two curves.}
\end{figure}

From Eq.~(\ref{eq:pol_dphi}) one sees that the polarization $p$ depends 
linearly on the geometrical factor $\gamma$.
Instead of measuring $p$ for a given value of $\gamma$, one can alternatively
also measure $\gamma$ for a given $p$. This grants access to another
physical variable of the topological insulator: the spin-flip mean
free path, which is crucial to applications in spintronics. The spin-flip
mean free path $\xi$ is the average path after which an electron
has lost information on its original spin and hence also on its
propagation direction. If there is spin-scattering
in the TI, the measured apparent spin-polarization of the surface
states will depend on the length of the path between the Al and M
electrode and the spin-flip mean free path $\xi$ thereby enters into
the distance dependent $\gamma$ factor. For the device in Fig.~\ref{fig:device}
we derived the probability distribution $f\left(\varphi\right)$ accounting for a finite spin-flip mean free
path $\xi$ in Appendix~\ref{sub:Common-spin-flip}. From this expression
$\gamma$ can be calculated numerically. For a known polarization
$p$ one can then simply calculate $\xi$ by fitting $\gamma$ to Eq.~\eqref{eq:pol_dphi}.
If $p$ is unknown, it is still possible to calculate $\xi$ from how
$\gamma$ varies with distance $d$. However, this is more inaccurate
as it requires multiple devices with different distances $d$. To reduce errors, the different devices
should at least be prepared on the same sample, since $\xi$ and $p$
may depend on the quality of the TI material. 
When we define the experimentally accessible quantity
\begin{equation}
\chi=\frac{\left(g_{1}-g_{3}\right)-\left(g_{4}-g_{2}\right)}{\left(g_{1}-g_{3}\right)+\left(g_{4}-g_{2}\right)}\frac{1}{\cos\Delta\varphi_{\textrm{Al}}},\label{eq:chi}
\end{equation}
$p$ is of the form
\begin{equation}
p=\chi_{i}\left(d_{i}\right)\gamma_{i}\left(d_{i}\right)\label{eq:p_chi}
\end{equation}
and identical for devices with different distances $d_{i}$. Then, we
can solve for the ratio of two $\gamma_{i}$:
\begin{equation}
\frac{\gamma_{1}\left(d_{1}\right)}{\gamma_{2}\left(d_{2}\right)}=\frac{\chi_{2}\left(d_{2}\right)}{\chi_{1}\left(d_{1}\right)}\;
.
\end{equation}
The spin-flip mean free path $\xi$ can be fitted to this 
ratio as shown in Fig.~\ref{fig:gamma_spin-flip}.
In order to get an accurate result, $d$ or, in the case of an unknown
polarization, the distances $d_{1}$, $d_{2}$ and $\left|d_{2}-d_{1}\right|$ should
be in the same order of magnitude as $\xi$.

\section{Summary and Conclusions}

We have studied a generalization of the Meservey-Tedrow method to
topological insulators.
Starting from an analytical approximation of TI surface states, we
showed how quantum tunneling from a superconductor into these surface
states can be used to measure their spin polarization. In contrast to
the application to ferromagnets, one has to measure the tunneling
current with respect to different spatial directions and take into
account that the density of states in TI surface states is strongly energy dependent.
Considering these aspects we derived formulas that allow easy calculation
of the in-plane spin polarization from measured tunneling spectra,
where the geometry of the device enters as a single factor. As there
is no chemical potential dependence of the in-plane polarization,
the tunneling spectra seem to be insensitive to the out-of-plane polarization
of the surface states, at least for the device geometries considered here.
When spin-flip scattering is included in the calculation of the geometrical
factor, it can be measured as well and hence, if measured prior to
the spin polarization, can increase the accuracy of the calculated
spin polarization.

\section*{Acknowledgments}

Financial support from the DFG via SPP 1666 ``Topological Insulators'' is gratefully acknowledged. 
We would like to thank A. Thomas and G. Reiss for valuable discussions.

\appendix

\section{\label{sec:Geometrical-factor}Geometrical factor}

The geometrical factor $\gamma$ plays a crucial role when calculating
the spin polarization from a given DC curve. In this appendix we will
therefore discuss the calculation of $\gamma$ for some alternative
geometries of the device. These geometries have some advantages over
the basic geometry in Fig.~\ref{fig:device}, but are more difficult
to realize. Considering the basic geometry in Fig.~\ref{fig:device}
we will also give a simple example of including the effect of spin-flip
scattering in the TI. The function $f\left(\varphi\right)$, from
which $\gamma$ is derived, gives the probability that an electron
starting at an angle $\varphi$ in the TI ends up at the considered
metallic electrode. For a given value of $\varphi$ it is hence given
by the ratio of the number of all possible trajectories reaching the
electrode and those not reaching it.

\subsection{Basic geometry with spin-flip scattering\label{sub:Common-spin-flip}}

In the basic geometry in Fig.~\ref{fig:device} it is assumed that
all electrons initially moving in positive $x$-direction will end
up at the electrode at $x>0$, which is only true as long as there
is no spin-flip scattering. If $\xi$ is the spin-flip mean-free-path
of the surface states, $d$ the distance between the aluminum and
the metallic electrode and $l$ the width of these electrodes, $f\left(\varphi\right)$
is given by 
\begin{eqnarray}
f\left(\varphi\right) & = & \int_{d}^{d+2l}dx\frac{l-\left|x-d-l\right|}{l^{2}}\\
 &  & \cdot\frac{1}{2}\begin{cases}
\left(1+e^{-\frac{x}{\xi\cos\varphi}}\right) & \textrm{if }\varphi\in\left[-\frac{\pi}{2},\frac{\pi}{2}\right]\\
\left(1-e^{\frac{x}{\xi\cos\varphi}}\right) & \textrm{if }\varphi\in\left[-\pi,-\frac{\pi}{2}\right],\left[\frac{\pi}{2},\pi\right]
\end{cases}\nonumber 
\end{eqnarray}
Here, $d\leq x\leq d+2l$ is the distance between two vertical lines
in the two electrodes and the integral averages over all $x$. Note,
that with spin-flip scattering also electrons initially moving in
the opposite direction can reach the electrode. For small $\xi$ or
large $x$, $f\left(\varphi\right)$ approaches $\frac{1}{2}$ for
all $\varphi$, i.e. all information on the initial spin is lost.
In order to measure the spin from such a device the ratio $\frac{x}{\xi}$
should therefore be small. The geometrical factor $\gamma$ has to
be calculated numerically for specific values of $\xi$, $d$ and
$l$ (see Fig.~\ref{fig:gamma_spin-flip}).

\subsection{Semi circles}

\begin{figure}
\centering\includegraphics[width=0.9\columnwidth]{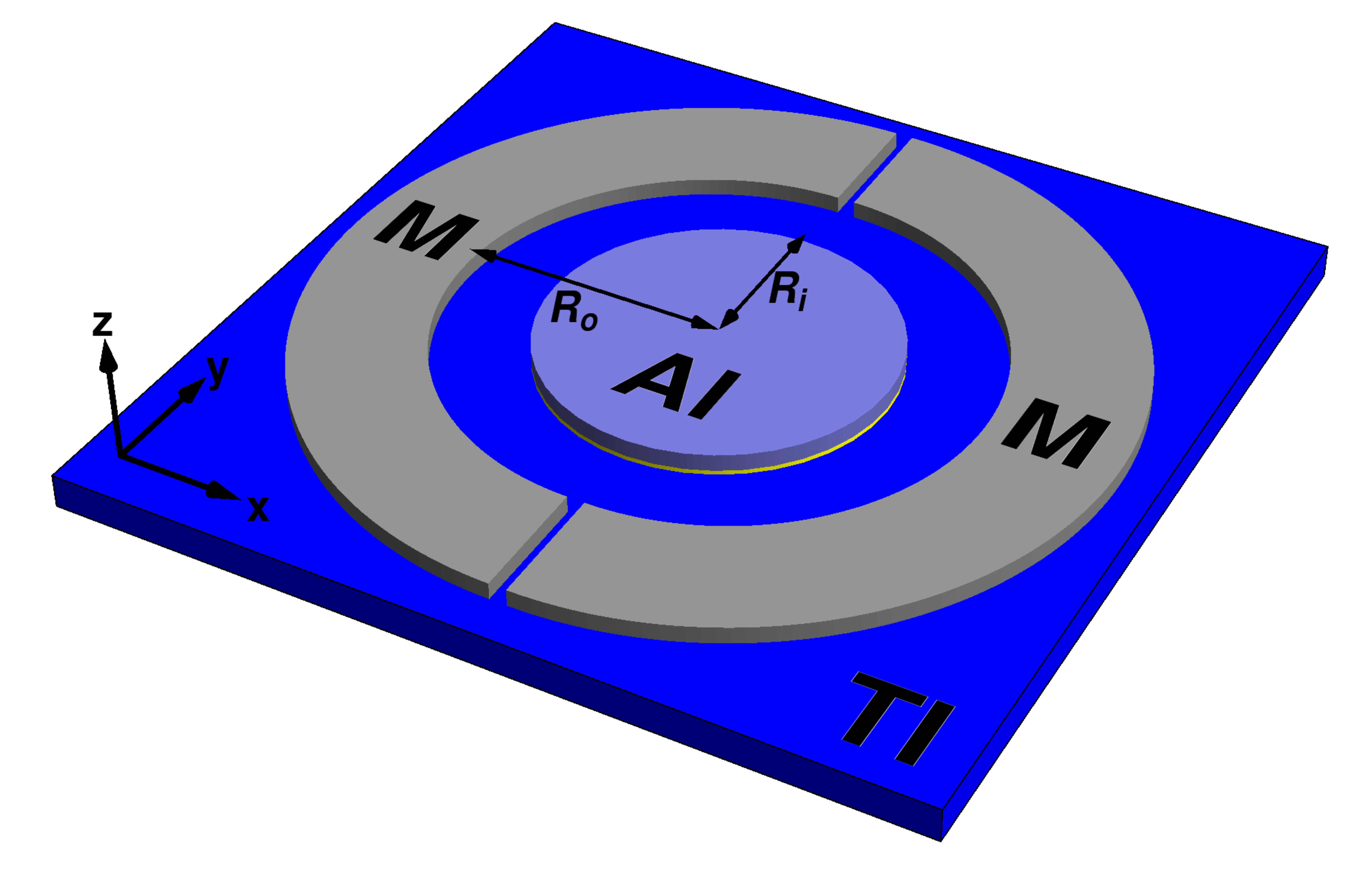}

\caption{\label{fig:Semi-circle-setup}Semi circle setup of the device. In
this setup, the mean path length for electrons in the TI is independent
of $\varphi$.}

\end{figure}

The concept of electrodes with a semi circle form (Fig.~\ref{fig:Semi-circle-setup})
can be interesting if the spin-flip mean-free-path of the surface
states is small, because in this geometry, the path length for electrons
in the TI is small in all directions, reducing spin-flip processes.
It also completely avoids possible scattering processes at the edges
of the TI. For a given angle $\varphi$ only electrons from a circular
segment of the round Al electrode can reach the electrode in positive
$x$-direction and $f\left(\varphi\right)$ is then given by the ratio
of this segment and the total area of the Al electrode: 
\begin{eqnarray}
f\left(\varphi\right) & = & 1-\frac{1}{\pi}\Re\left[\arccos\left(\frac{R_{o}}{R_{i}}\cos\varphi\right)\right.\\
 &  & \left.-\frac{R_{o}}{R_{i}}\cos\varphi\sqrt{1-\frac{R_{o}^{2}}{R_{i}^{2}}\cos^{2}\varphi}\right]\nonumber \\
 & \overset{\left(R_{o}\approx R_{i}\right)}{\approx} & 1-\frac{1}{\pi}\left(\left|\varphi\right|-\cos\varphi\sin\left|\varphi\right|\right)
\end{eqnarray}
Here, $R_{i}$ and $R_{o}$ are the inner and outer radii of the spacing
between aluminum (Al) and metallic (M) electrode, with $R_{o}-R_{i}$
being the distance between them. In the limit $R_{o}\approx R_{i}$ we
get $\gamma=\frac{3}{16}\pi^{2}$, while for $\frac{R_{i}}{R_{o}}\rightarrow0$
the solution $\gamma=\frac{\pi}{2}$ of the basic geometry is recovered.

\subsection{U-shaped geometry\label{sub:U-shaped-geometry}}

\begin{figure}
\centering\includegraphics[width=0.9\columnwidth]{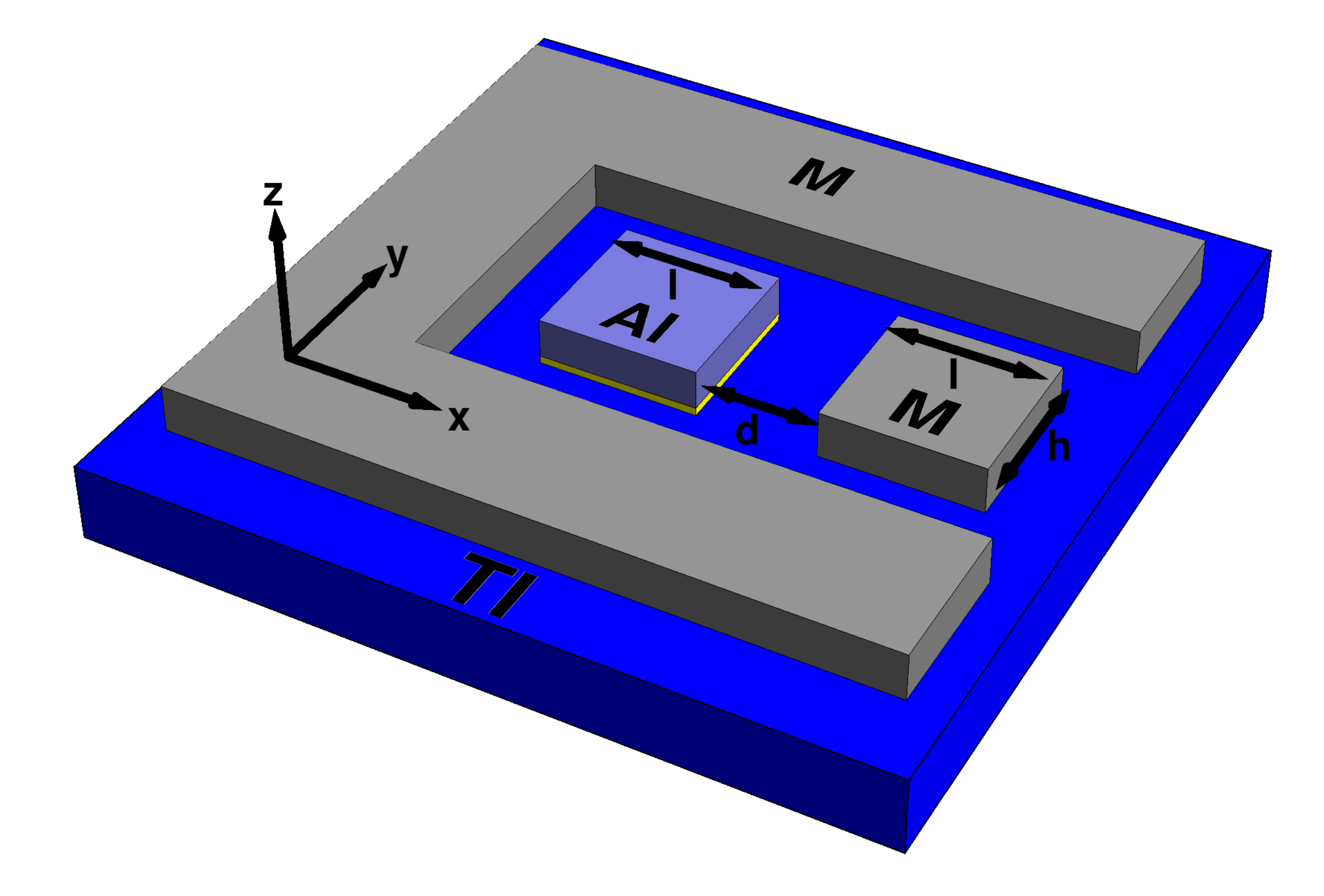}

\caption{\label{fig:U-shaped-setup}U-shaped setup of the device (modified
from Ref.~\onlinecite{TMR}). The two relevant electrodes which
are separated by $d$ are assumed to have both the same height $h$
and length $l$. The U-shaped electrode captures nearly all electrons
moving in other directions than the small metallic electrode.}

\end{figure}

The device with a U-shaped metallic electrode (Fig.~\ref{fig:U-shaped-setup})
discussed in Ref.~\onlinecite{TMR} opens the possibility to investigate
the spin-polarization of surface states with a specific momentum,
because it filters out almost all electrons moving in other directions
than that given by the two small electrodes. Hence, one could even
apply it to anisotropic surfaces, where the Fermi surface is not circular
and the spin-polarization may depend on the momentum direction. In
this case $f\left(\varphi\right)$ is given by\cite{TMR}
\begin{eqnarray}
f\left(\varphi\right) & = & \int_{d}^{d+2l}dx\frac{l-\left|x-d-l\right|}{l^{2}}\label{eq:U-shaped}\\
 &  & \cdot\left(1-\frac{x}{h}\tan\left|\varphi\right|\right)\Theta\left(\arctan\frac{h}{x}-\left|\varphi\right|\right),\nonumber 
\end{eqnarray}
where $\Theta$ is the Heaviside step function, $l$ and $h$ are
the length and height of the two small electrodes and $d$ the distance
between them. For given values of $d$, $l$ and $h$, $\gamma$ can
be calculated numerically. If $\frac{x}{h}$ is large for all $x$,
Eq.~\eqref{eq:U-shaped} may be approximated by $f\left(\varphi\right)=\delta\left(\varphi\right)$,
i.e. $\gamma=1$. The DC then effectively becomes that of a two dimensional
device, where all electrons reaching the metallic electrode have the
same momentum direction and spin. In principle, using multiple devices with different
orientations of the electrodes with respect to the lattice of the
TI the $\mathbf{k}$-dependent spin structure of the surface states
can be studied. It is also possible to split the U-shaped electrode
into multiple parts. In this way measurements in multiple directions can be
done with a single device.


\begin{thebibliography}{10}
\bibitem{Bernevig} B.~A.~Bernevig, T.~L.~Hughes, and S.-C.~Zhang,
Quantum Spin Hall Effect and Topological Phase Transition in HgTe
Quantum Wells, \href{http://dx.doi.org/10.1126/science.1133734}{Science {\bf 314}, 1757 (2006)}.

\bibitem{Fu} L. Fu, C.~L. Kane and E.~J. Mele, Topological Insulators
in Three Dimensions, \href{http://dx.doi.org/10.1103/PhysRevLett.98.106803}{Phys. Rev. Lett. {\bf 98}, 106803 (2007)}.

\bibitem{Koenig} M.~K\"onig, S.~Wiedmann, C.~Br\"une, A.~Roth, H.~Buhmann,
L.W.~Molenkamp, X.-L.~Qi, and S.-C.~Zhang, Quantum Spin Hall Insulator
State in HgTe Quantum Wells \href{http://dx.doi.org/10.1126/science.1148047}{Science {\bf 318}, 766 (2007)}.

\bibitem{Hsieh1} D. Hsieh, D. Qian, L. Wray, Y. Xia, Y. Hor, R.J.
Cava, and M.Z. Hasan, A topological Dirac insulator in a quantum spin
Hall phase, \href{http://dx.doi.org/10.1038/nature06843}{Nature (London) {\bf 452}, 970 (2008)}.

\bibitem{Chen}Y.~L.~Chen, J.~G.~Analytis, J.-H.~Chu, Z.~K.~Liu,
S.-K.~Mo, X.~L.~Qi, H.~J.~Zhang, D.~H.~Lu, X.~Dai, Z.~Fang,
S.~C.~Zhang, I.~R.~Fisher, Z.~Hussain, and Z.-X.~Shen, Experimental
Realization of a Three-Dimensional Topological Insulator, Bi$_{2}$Te$_{3}$,
\href{http://dx.doi.org/10.1126/science.1173034}{Science {\bf 325}, 178 (2009)}.

\bibitem{Tanaka} Y. Tanaka, T. Yokoyama, and N. Nagaosa, Manipulation
of the Majorana Fermion, Andreev Reflection, and Josephson Current
on Topological Insulators, \href{http://link.aps.org/doi/10.1103/PhysRevLett.103.107002}{Phys. Rev. Lett. {\bf   103}, 107002 (2009)}.

\bibitem{Garate} I. Garate and M. Franz, Inverse Spin-Galvanic Effect
in the Interface between a Topological Insulator and a Ferromagnet,
\href{http://link.aps.org/doi/10.1103/PhysRevLett.104.146802}{Phys. Rev. Lett. {\bf 104}, 146802 (2010)}.

\bibitem{Krueckl} V.~Krueckl and K.~Richter, Switching Spin and
Charge between Edge States in Topological Insulator Constrictions,
\href{http://link.aps.org/doi/10.1103/PhysRevLett.107.086803}{Phys. Rev. Lett. {\bf 107}, 086803 (2011)}.

\bibitem{TMR}M.~G\"otte, T.~Paananen, G.~Reiss, and T.~Dahm, Tunneling
Magnetoresistance Devices Based on Topological Insulators: Ferromagnet\textendash Insulator\textendash Topological-Insulator
Junctions Employing Bi$_{2}$Se$_{3}$, \href{http://dx.doi.org/10.1103/PhysRevApplied.2.054010}{Phys. Rev. Applied {\bf 2}, 054010 (2014)}.

\bibitem{GJD}M. G\"otte, M. Joppe, and T. Dahm, Pure spin current devices
based on ferromagnetic topological insulators,
\href{http://doi.org/10.1038/srep36070}{Sci. Rep. {\bf 6}, 36070 (2016)}.

\bibitem{Han}J. Han, A. Richardella, S.A. Siddiqui, J. Finley, N. Samarth, and
  L. Liu, Room-Temperature Spin-Orbit Torque Switching Induced by a
  Topological Insulator,
  \href{http://link.aps.org/doi/10.1103/PhysRevLett.119.077702}{Phys. Rev. Lett. {\bf
      119}, 077702 (2017)}.

\bibitem{Khang}N.~H.~D.~Khang, Y.~Ueda, and P.~N.~Hai, A conductive
  topological insulator with large spin Hall effect for ultralow power
  spin-orbit torque switching,
  \href{https://doi.org/10.1038/s41563-018-0137-y}{Nat. Mat. {\bf 17}, 808
  (2018)}.

\bibitem{He}M.~He, H.~Sun, and Q.~L.~He, Topological Insulator: Spintronics
  and Quantum Computation,
  \href{https://doi.org/10.1007/s11467-019-0893-4}{Front. Phys. {\bf 14},
    43401 (2019)}

\bibitem{Liu:PRB10} C.-X.~Liu, X.-L.~Qi, H.J.~Zhang, X.~Dai,
Z.~Fang, and S.-C.~Zhang, Model Hamiltonian for topological insulators,
\href{http://link.aps.org/doi/10.1103/PhysRevB.82.045122}{Phys. Rev. B {\bf 82}, 045122 (2010)}.

\bibitem{Souma}S. Souma, K. Kosaka, T. Sato, M. Komatsu, and A. Takayama,
Direct Measurement of the Out-of-Plane Spin Texture in the Dirac-Cone
Surface State of a Topological Insulator, \href{http://dx.doi.org/10.1103/PhysRevLett.106.216803}{Phys. Rev. Lett. {\bf 106}, 216803 (2011)}.

\bibitem{Yazyev}O.~V.~Yazyev, J.~E.~Moore, and S.~G.~Louie,
Spin Polarization and Transport of Surface States in the Topological
Insulators Bi$_{2}$Se$_{3}$ and Bi$_{2}$Te$_{3}$ from First Principles,
\href{http://dx.doi.org/10.1103/PhysRevLett.105.266806}{Phys. Rev. Lett. {\bf 105}, 266806 (2010)}.

\bibitem{Wang}X.~Wang, G.~Bian,~T.~Miller, and T.-C.~Chiang,
Topological spin-polarized electron layer above the surface of Ca-terminated
Bi$_{2}$Se$_{3}$, \href{http://dx.doi.org/10.1103/PhysRevB.87.035109}{Phys. Rev. B {\bf 87}, 035109 (2013)}.

\bibitem{Sanchez} J.~Sanchez-Barriga et al, Photoemission of Bi$_{2}$Se$_{3}$
with Circularly Polarized Light: Probe of Spin Polarization or Means
for Spin Manipulation?, \href{http://dx.doi.org/10.1103/PhysRevX.4.011046}{Phys. Rev. X {\bf 4}, 011046 (2014)}.

\bibitem{Pan:PRL106} Z.-H.~Pan, E.~Vescovo, A.~V.~Fedorov, D.~Gardner,
Y.~S.~Lee, S.~Chu, G.~D.~Gu, and T.~Valla, Electronic Structure
of the Topological Insulator Bi$_{2}$Se$_{3}$ Using Angle-Resolved
Photoemission Spectroscopy: Evidence for a Nearly Full Surface Spin
Polarization, \href{http://link.aps.org/doi/10.1103/PhysRevLett.106.257004}{Phys. Rev. Lett. {\bf 106}, 257004 (2011)}.

\bibitem{Jozwiak}C.~Jozwiak, Y.~L.~Chen, A.~V.~Fedorov, J.~G.~Analytis,
C.~R.~Rotundu, A.~K.~Schmid, J.~D.~Denlinger, Y.-D.~Chuang,
D.-H.~Lee, I.~R.~Fisher, R.~J.~Birgeneau, Z.-X.~Shen, Z.~Hussain,
and A.~Lanzara, Widespread spin polarization effects in photoemission
from topological insulators, \href{http://dx.doi.org/10.1103/PhysRevB.84.165113 }{Phys. Rev. B {\bf 84}, 165113 (2011)}.

\bibitem{Landolt}G.~Landolt, S.~Schreyeck, S.~V.~Eremeev, B.~Slomski,
S.~Muff, J.~Osterwalder, E.~V.~Chulkov, C.~Gould, G.~Karczewski,
K.~Brunner, H.~Buhmann, L.~W.~Molenkamp, and J.~H.~Dil, Spin
Texture of Bi$_{2}$Se$_{3}$ Thin Films in the Quantum Tunneling
Limit,\href{http://dx.doi.org/10.1103/PhysRevLett.112.057601}{Phys. Rev. Lett. {\bf 112}, 057601 (2014)}.

\bibitem{Park}C.-H. Park and S. G. Louie, Spin Polarization of Photoelectrons
from Topological Insulators, \href{http://dx.doi.org/10.1103/PhysRevLett.109.097601}{Phys. Rev.Lett. {\bf 109}, 097601 (2012)}.

\bibitem{TedrowMeservey-PRL26}R.~Meservey and P.~M.~Tedrow, Spin-Dependent
Tunneling into Ferromagnetic Nickel, \href{http://dx.doi.org/10.1103/PhysRevLett.26.192}{Phys. Rev. Lett. {\bf 26}, 192 (1971)}.

\bibitem{TedrowMeservey-PRB7}R.~Meservey and P.~M.~Tedrow, Spin
Polarization of Electrons Tunneling from Films of Fe, Co, Ni, and
Gd, \href{http://dx.doi.org/10.1103/PhysRevB.7.318}{Phys. Rev. B {\bf 7}, 318 (1973)}.

\bibitem{TedrowMeservey-PRL25}R.~Meservey, P.~M.~Tedrow and P.~Fulde,
Magnetic Field Splitting of the Quasiparticle States in Superconducting
Aluminum Films, \href{http://dx.doi.org/10.1103/PhysRevLett.25.1270}{Phys. Rev. Lett. {\bf 25}, 1270 (1970)}.

\bibitem{Dissertation}M. G\"otte, Topological insulator based spintronics: Theoretical investigation of pure spin current devices and polarization measurements, Universit\"at Bielefeld, Bielefeld (2017), \href{https://nbn-resolving.de/urn:nbn:de:0070-pub-29128707}{urn:nbn:de:0070-pub-29128707}

\bibitem{Dufouleur}J.~Dufouleur, L- Veyrat, B.~Dassonneville, C.~Nowka,
  S.~Hampel, P.~Leksin, B.~Eichler, O.~G.~Schmidt, B.~B\"uchner, and
  R.~Giraud, Enhanced Mobility of Spin-Helical Dirac Fermions in Disordered 3D
  Topological Insulators, 
\href{https://doi.org/10.1021/acs.nanolett.6b02060}{Nano Lett. {\bf 16}, 6733 (2016)}.

\bibitem{Koenig2} M.~K\"onig, H.~Buhmann, L.~W.~Molenkamp, T.~Hughes,
C.-X.~Liu, X.-L.~Qi, and S.-C.~Zhang, The Quantum Spin Hall Effect: Theory and Experiment
\href{http://dx.doi.org/10.1143/JPSJ.77.031007}{J. Phys. Soc. Japan {\bf 77}, 031007 (2008)}.

\bibitem{Hao}L.~Hao and T.~K.~Lee, Surface spectral function in
the superconducting state of a topological insulator, \href{http://dx.doi.org/10.1103/PhysRevB.83.134516}{Phys. Rev. B {\bf 83}, 134516 (2011)}.

\bibitem{Zhang:APL09}G.~Zhang, H.~Qin, J.~Teng, J.~Guo, Q.~Guo,
X.~Dai, Z.~Fang, and K.~Wu, Quintuple-layer epitaxy of thin films
of topological insulator Bi$_{2}$Se$_{3}$,
\href{http://dx.doi.org/10.1063/1.3200237}{Appl. Phys. Lett. {\bf 95}, 053114
  (2009)}.

\bibitem{Dynes}R.~C.~Dynes, V.~Narayanamurti, and J.~P.~Garno, Direct
  Measurement of Quasiparticle-Lifetime Broadening in a Strong-Coupled
  Superconductor, \href{http://dx.doi.org/10.1103/PhysRevLett.41.1509}{Phys. Rev. Lett. {\bf 41}, 1509 (1978)}.

\bibitem{Schebaum}O.~Schebaum, D.~Ebke, A.~Niemeyer, G.~Reiss, J.~S.~Moodera, A.~Thomas, Direct measurement of the spin polarization of Co$_2$FeAl in combination with MgO tunnel barriers, \href{https://doi.org/10.1063/1.3358245}{J. Appl. Phys. 107, 09C717 (2010)}.

\end{thebibliography}
\end{document}